\documentclass[fleqn,usenatbib]{mnras}

\usepackage{pdflscape}
\usepackage{rotating}
\usepackage{tabularray}

\usepackage{graphicx}
\usepackage{txfonts}
\usepackage{natbib,twoopt}
\usepackage{upgreek}

\graphicspath{{./}{fig/}}

\newcommand{\IUE}{{\it IUE}}
\newcommand{\FUSE}{{\it FUSE}}
\newcommand{\HST}{{\it HST}}

\usepackage[T1]{fontenc}

\DeclareRobustCommand{\VAN}[3]{#2}
\let\VANthebibliography\thebibliography
\def\thebibliography{\DeclareRobustCommand{\VAN}[3]{##3}\VANthebibliography}

\usepackage{graphicx}	

\title[Disc spectra in DNe and NLs]{Revisiting the accretion disc spectra of Dwarf Novae and Novalike variables: implications for the standard disc model}

\author[G. Zsidi et al.]{
Gabriella Zsidi,$^{1, 2, 3}$\thanks{E-mail: g.zsidi@leeds.ac.uk}
C. J. Nixon,$^{1}$
T. Naylor$^{4}$
and J. E. Pringle$^{5}$
\\
$^{1}$School of Physics and Astronomy, Sir William Henry Bragg Building, Woodhouse Ln.,
    University of Leeds, Leeds LS2 9JT, UK\\
$^{2}$Konkoly Observatory, HUN-REN Research Centre for Astronomy and Earth Sciences, Konkoly-Thege Miklós út 15-17, 1121 Budapest, Hungary\\
$^{3}$CSFK, MTA Centre of Excellence, Budapest, Konkoly Thege Mikl\'os \'ut 15-17, 1121, Hungary\\
$^{4}$Department of Physics and Astronomy, University of Exeter, Stocker Road, Exeter EX4 4QL, UK\\
$^{5}$Institute of Astronomy, University of Cambridge, Madingley Road, Cambridge, CB3 0HA, UK
}

\date{Accepted XXX. Received YYY; in original form ZZZ}

\pubyear{2024}

\begin{document}
\label{firstpage}
\pagerange{\pageref{firstpage}--\pageref{lastpage}}
\maketitle

\begin{abstract}
Accretion discs are fundamental to much of astronomy. They can occur around stars both young and old, around compact objects they provide a window into the extremes of physics, and around supermassive black holes in galaxy centres they generate spectacular luminosities that can outshine the entire galaxy. However, our understanding of the inner workings of accretion discs remains far from complete. Here we revisit a conundrum in the observations of some of the simplest accreting systems; the Cataclysmic Variables (CVs).  The high-accretion-rate states of (non-magnetic) CVs can be divided into the short-lived outbursts ($\sim$ a week) typical of dwarf novae (DNe) and the long-lived (and sometimes perpetual) high states of nova-like (NL) CVs.  Since both sorts of high-state occur in approximately steady-state accretion discs with similar properties and accretors, we would expect them to display similar spectral energy distributions. However, previous analyses based on UV spectra from the {\it International Ultraviolet Explorer} have shown that their spectral energy distributions are different. We perform a re-analysis of the data using up to date calibrations and distance (and thus dereddening) estimates to test whether this difference persists and whether it is statistically significant over the sample. We find that it does persist and it is statistically significant. We propose routes to investigating this discrepancy further and discuss the implications this has for other accreting systems, such as X-ray binaries, Active Galactic Nuclei and protoplanetary discs.
\end{abstract}




\section{Introduction}
Accretion discs occur in many astrophysical systems as a consequence of the conservation of angular momentum in material that is falling due to the gravitational pull of an accretor \citep{lyndenbell1969,pringle1972,shakura1973,pringle1981}. Discs form around protostars in star forming regions \citep[e.g.][]{armitage2011}, around main-sequence stars \citep[e.g.][]{rivinius2013}, due to mass transfer in binary systems \citep[e.g.][]{warner1995} and around the supermassive black holes in galaxy centres \citep[AGN;][]{lyndenbell1969}.

The basics of the standard model of accretion discs are laid out in \cite{pringle1972} and \cite{shakura1973}, and summarised in \cite{pringle1981}. 
The disc matter orbits the central object in a sequence of planar and circular orbits until arriving at the inner boundary, the location of which is set by the physics of the central object and, in some cases, the properties of the disc itself. At each radius in the disc, the material must give up some orbital energy to move to a smaller radius. 
This is effected by viscous torques within the disc that cause angular momentum to be transferred outwards and mass inwards, and the net energy transfer results in heating of the disc material. 
This heat is radiated from the disc surface and this is what we see from the accretion disc. 

A basic assumption in the standard model is that the accretion energy is radiated locally from the disc in the region in which it is produced. If this is the case, and assuming a steady supply of mass from large radii and that the disc surface radiates as a blackbody, the disc temperature profile takes a simple form that is independent of the details of the viscous process that produces the energy dissipation. The temperature profile is \citep[e.g.][]{pringle1981}
\begin{equation}
T(R) = \left\{\frac{3GM{\dot M}}{8\pi R^3\sigma_{\rm SB}}\left[1-\left(\frac{R_{\rm in}}{R}\right)^{1/2}\right]\right\}^{1/4}
\end{equation}
where ${\dot M}$ is the accretion rate on to the central object of mass $M$ and radius $R_{\rm in}$, and $\sigma_{\rm SB}$ is the Stefan-Boltzmann constant. 
An implication of this simple equation is that, for a central accretor of the same mass and radius accreting at the same rate, we should expect to see basically the same spectrum emitted from the disc. 

In many cases the environment and dynamics of accretion flows are complex and this can make direct comparisons between models and observational data difficult. 
However, as our understanding of the inner workings of accretion discs is not yet complete, it is important to test the basic assumptions that are made in disc models where we can. 
To keep things as simple as possible, we need to find systems where the accretion flow is approximately time-steady (at least in the inner regions where most of the energy is produced), with simple central objects, and surface temperatures in the regime where we have a good understanding of the emitted spectra. 
These properties are satisfied by some cataclysmic variable (CV) stars, particularly the dwarf novae (DNe) and the novalike variables (NLs).\footnote{Here we restrict our attention to the ``non-magnetic'' cases, where there is no evidence that the central object hosts a sufficiently strong magnetic field as to interfere with the accretion flow.}

CVs are close binaries, in which material is transferred from a Roche-lobe filling secondary to a white dwarf primary through the L1 point \citep{warner1995}. 
For non-magnetic CVs, where the magnetic field strength of the white dwarf is insufficient to significantly perturb the accretion flow, the transferred material forms an accretion disc that extends down to the surface of the white dwarf \citep{lubow1975}. 
The subclass of CVs known as DNe undergo a thermal-viscous instability cycle, during which the accretion disc displays two different accretion regimes: (1) quiescent states in which the disc viscosity parameter $\alpha$ is low and mass accumulates in the disc over time and (2) outburst states in which the disc viscosity parameter is high and mass drains on to the central white dwarf \citep[e.g.][]{meyer1981, faulkner1983, smak1984, warner1995, lasota2001}.\footnote{In the quiescent states we do not have a good handle on the magnitude of the \cite{shakura1973} viscosity parameter $\alpha$ with estimates of $\alpha \lesssim 0.01$. 
In the outburst states $\alpha$ is measured to be around 0.3 (e.g. \citealt{kotko2012}; see the summaries in \citealt{king2007}; \citealt{martin2019}).} 
During the outbursts, which typically last for around a week, the disc dominates the electromagnetic output of the system. NLs do not exhibit DNe outbursts,\footnote{They can exhibit brightening events or flares, but these are mostly attributed to mass transfer variations from the companion star (see, e.g., \citealt{zsidi2023} and references therein).} and instead exhibit luminosities that are commensurate with the outburst states of the DNe for extended periods of time. 
The accepted explanation for the difference between these two sub-classes of CV is that the NLs have a high enough mass transfer rate from the companion star to exist permanently in the high state, whereas the DNe, having smaller mass transfer rates, must suffer the limit cycle behaviour induced by the thermal-viscous instability. 
Naively one may therefore expect the NLs to appear bluer than the DNe as the disc temperature grows slightly with increasing ${\dot M}$. 
But it is also possible that, as the NLs can be intrinsically brighter, we may typically observe them at larger distances making them more vulnerable to interstellar reddening.

To explore the differences between the DN and NL classes, \cite{ladous1991} undertook a statistical analysis of the ultraviolet (UV) spectra of 32 DNe and 23 NLs using data from the {\it International Ultraviolet Explorer} (\IUE) satellite, and found that the NLs appear to be somewhat redder\footnote{By ``redder'' we mean a larger ratio of long-wavelength to short-wavelength flux.} than the DNe in outburst. 
More recently, \cite{hamilton2007} fit the \IUE\ DNe outburst spectra with the synthetic accretion disc spectra of \cite{wade1998}, finding that the synthetic spectra provide a good fit for almost all of the DNe outburst spectra. 
From this we can surmise that the standard disc model is a good description of the accretion process in DNe. 
However, the same cannot be said for NLs. 
Following initial studies on individual systems by \cite{wade1984,wade1988}, \cite{ladous1991} used a large sample of CVs to show that \IUE\ spectra of NLs are redder than the DNe outburst spectra, and thus are not well fit by the synthetic models \citep[see also][]{puebla2007}.
Subsequent models that provide a good fit with NL spectra (from \IUE\ and also \FUSE\ and \HST) are provided by \cite{long1994}, \cite{linnell2007,linnell2007b}, \cite{linnell2010} and \cite{godon2017}. 
These models artificially increase the inner radius of the disc to several times the size of the white dwarf; this has the effect of reducing the peak temperature of the inner disc, making the spectrum redder as a result (some models artificially cap the temperature at a maximum value or change the radial power-law of the temperature profile instead of moving the inner boundary). 
In agreement with \cite{ladous1991}, \cite{godon2017} explore the disc-dominated \IUE\ spectra of CVs and find that NL spectra are typically redder than the DNe in outburst. 

In this paper, we aim to build on the analyses presented by \cite{ladous1991} and \cite{godon2017}, by performing a reanalysis of the \IUE\ spectra taking account of updated calibrations of the \IUE\ pipeline, updated distances that allow for more accurate dereddening of the spectra, and to perform statistical tests of the samples to firmly establish the significance of the difference in the colour of the \IUE\ spectra of DNe and NLs. 
In Section~\ref{sect:observations} we discuss the \IUE\ data, sample selection and dereddening calculations. In Section~\ref{sect:results} we present our analysis of the colours of DNe and NLs. In Section~\ref{sect:discussion} we present discussion of the results and we conclude and provide implications for other systems in Section~\ref{sect:conclusion}.

\section{Archival {\it IUE} observations}
\label{sect:observations}
In this work we use data from \IUE, which carried out observations between 1978 January and 1996 September. 
The spacecraft was equipped with a short-wavelength prime (SWP: $1150-2000\,\AA$), a long-wavelength prime (LWP: $1850-3400\,\AA$) and a long-wavelength redundant camera (LWR: $1850-3400\,\AA$), and could provide both high- and low-dispersion spectra with a resolution of 0.2\,$\AA$ and 6\,$\AA$. 
In practice, the LWR provided better quality data and thus was generally used in preference to the LWP. 
To maintain a consistent dataset we consider only the low-dispersion spectra as these are more numerous.

The original spectroscopic image processing system, IUESIPS, was released in 1977, and it removed instrumental signatures and provided wavelength and flux calibration. 
This data reduction pipeline has, however, evolved throughout the years as users discovered certain shortcomings. 
This introduced inconsistencies to the \IUE\ data base \citep{nichols1996}. 
After \IUE\ was deactivated in 1996, the entire \IUE\ database was reprocessed in order to compile a high-quality and uniformly processed \IUE\ Final Archive. 
These reprocessed data products are called NEWSIPS \citep[New Spectroscopic Image Processing System;][]{nichols1994, nichols1996}. 
New image processing techniques were developed with this procedure, which included e.g. the improvement of the photometric calibration, new models for flux calibration, and also uniform wavelength calibration. 
The new procedure also resulted in an increase in the S/N of $10-50$ per cent for the low dispersion spectra. 
A detailed comparison between the IUESIPS and the NEWSIPS data products with example spectra can be found in \cite{nichols1996}.

\begin{figure}
    \includegraphics[width=0.99\columnwidth]{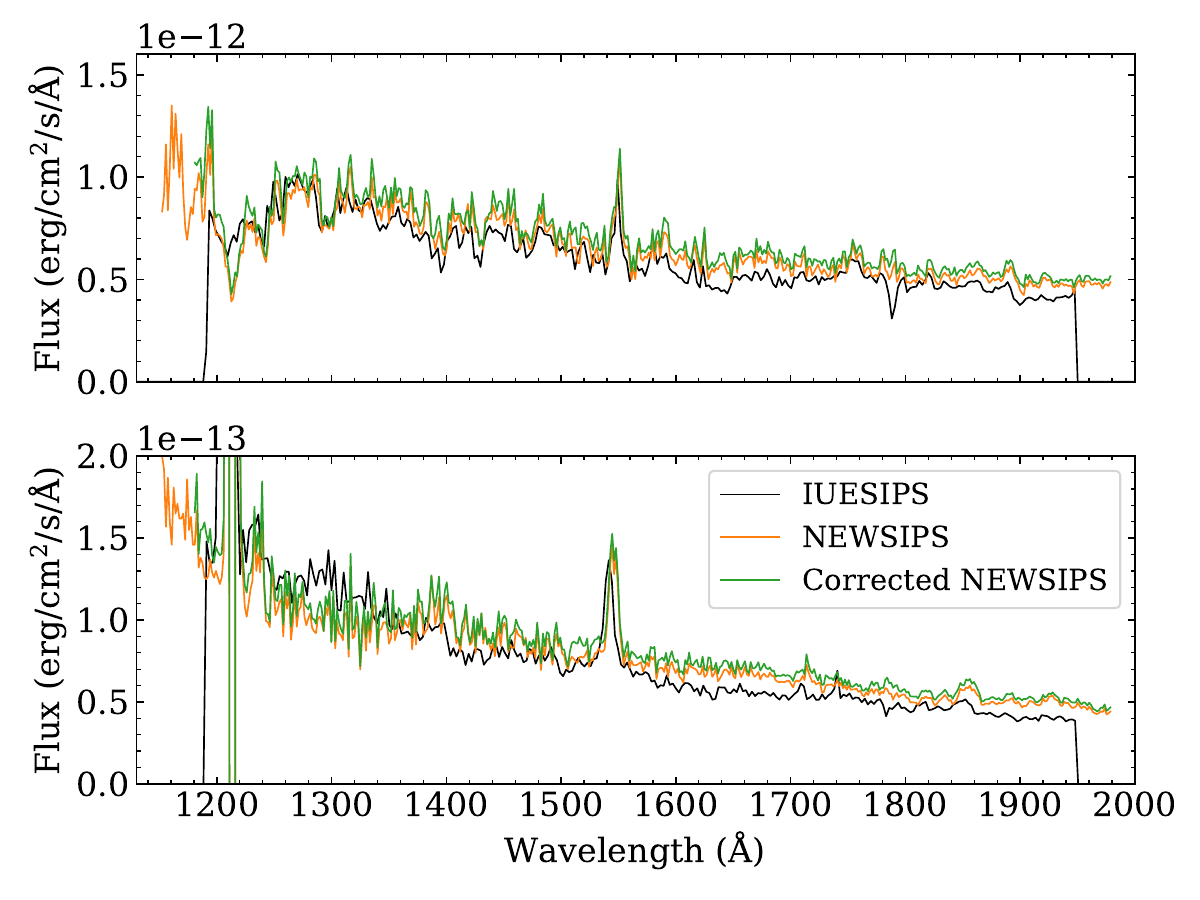}
    \caption{SWP10131 (top) and SWP10614 (bottom) spectra of TT~Ari. The black lines show the IUESIPS spectra, the orange lines show the NEWSIPS spectra, and the green lines show the NEWSIPS spectra after applying the \protect\cite{massa2000} correction. This shows that the UV spectral slope can be affected by the new \IUE\ data reduction pipeline (NEWSIPS) and, to a lesser extent, the \protect\cite{massa2000} correction.}
    \label{fig:ttari_spectra_compare}
\end{figure}

The NEWSIPS data showed significant improvement compared to the data products from the IUESIPS pipeline. 
However, comparison between \IUE\ low-dispersion data, models, and {\it Hubble Space Telescope} (\HST) observations revealed some problems with the absolute flux calibration, the thermal corrections, and time-dependent sensitivity \citep{massa2000}. 
In certain cases, this could result in discrepancies as large as 10\%. \cite{massa2000} provide the correction files and corresponding IDL routine.\footnote{\url{https://archive.stsci.edu/iue/contrib.html}} 
In Fig.~\ref{fig:ttari_spectra_compare}, we show an example spectrum of TT~Ari using the IUESIPS pipeline (employed by \citealt{ladous1991}), using the NEWSIPS data, and a ``Corrected NEWSIPS'' which include the \cite{massa2000} corrections.

The target acquisition of \IUE\ was carried out by the Fine Error Sensor (FES), which also recorded brightness information of the observed system. 
The raw FES counts are often available from the file headers for bright sources, which can be used to estimate the visual brightness of the system (FES magnitudes)\footnote{See section 4.7. of the \IUE\ observing guide for further details: \url{https://archive.stsci.edu/iue/instrument/obs_guide/}} using the calibration of \cite{perez1991}.
When using this calibration we ignore the colour corrections as they are small for DNe in outburst and NLs which have $B-V \approx 0$ \citep{warner1995}.


\subsection{Sample selection}
Our starting point was the sample presented in \cite{ladous1991}. 
\cite{ladous1991} studied the \IUE\ spectra obtained between 1978 and 1987, and her sample included 21 DNe and 11 NLs, whose inclination is $i \leq 60^{\circ}$ (when using the updated system parameters) and both short and long wavelength spectra were available. 
We collected updated system parameters for each object, which we list in Tables~\ref{tab:system_par_dne} and \ref{tab:system_par_nl}.
In order to restrict our attention to systems in which an accretion disc is present, clearly observable, and extends to the white dwarf surface, we only include non-magnetic systems and those with low inclination ($i \leq 60^{\circ}$).

 We discarded EK~TrA from our sample due to its unlikely large distance of 6650$^{+2820}_{-2199}$\,pc \citep{bailerjones2021}, and SW~UMa due to its likely magnetic nature \citep{shafter1986, dubus2018}. 
 We did not find contemporaneous photometric observations for V794~Aql, but its faint spectra suggest that it was probably in low state. 
 There were also no available light curve for V425~Cas from the American Association of Variable Star Observers (AAVSO) data base, but \cite{szkody1985} reports that the system was in a mid-state during a decline from maximum to minimum. 
 For these reasons, we removed V794~Aql and V425~Cas from our sample. 
 \IUE\ carried out further observations until 1997, which allows us to increase the number of objects in our sample. When selecting further sources from the \IUE\ archive, we followed the selection criteria by \cite{ladous1991}. 
 Namely, we cross-correlated the JD of the \IUE\ spectra with light curves in order to ensure that the sources we include are DNe in outburst and NLs in high state. 
 The AAVSO data base provides the most complete collection of light curves from the times of the \IUE\ observations but, unfortunately, these light curves do not always provide contemporaneous information on the source brightness. 
 We only included those new DNe, which we are confident are in outburst, and this gave us an additional 6 DNe. 
 The light curves of the NLs are very sparse, and we therefore attempted to use the FES counts to estimate the system brightness. 
 Many of these systems were too faint in the optical for the FES and the acquisition was made in `Blind Offset' mode, which means that no FES counts were recorded for the target. 
 Other objects were either in low state or are high inclination eclipsing systems, so our search for new NLs matching our criteria did not return any additional NLs. 
 Therefore, our total sample consists of 28 DNe and 8 NLs, as V426~Oph is classified as a DNe according to more recent catalogues \citep[e.g.][]{ritter2003}. 
 In all cases, we used the low dispersion large aperture observations, and we chose datasets where the typical time lag between the observations carried out by the short-wavelength (SWP) and the long-wavelength (LWP or LWR) camera is approximately half an hour (or in rare cases up to a few hours).

\subsection{Distances and Dereddening}\label{sec:dereddening}

We determined individual extinctions to each CV using the extinction maps of \cite{vergely2022} which are derived using measured extinctions and {\it Gaia} parallaxes of about 35 million stars. 
This required distances to each CV, for which we used the geometric distances given by \cite{bailerjones2021} based on the {\it Gaia} parallaxes. 
We do not use the photogeometric distances since the prior for these distances includes assumptions about the absolute magnitude of the star which are inappropriate for peculiar objects such as CVs. 
It is normal when using {\it Gaia} parallaxes to exclude stars where the astrometric fits to the individual observations are poor, expressed as a high renormalized unit weight error (RUWE, see for example, \citealt{Fabricius2021}). 
However we did not apply any such quality cut because it appears that CVs have anomalously poor summary statistics for the quality of their astrometric fits, which are not indicative of poor parallaxes \citep[see Section 2.1 of][]{Pala2020}. 
We suspect this is due to their variability.

Using these distances we obtained extinctions to each CV by integrating through the 10\,pc resolution \cite{vergely2022} data cubes from the Sun to the CV. 
The \cite{vergely2022} extinction cubes are created using the \cite{fitzpatrick2019} extinction law, which is qualitatively different from previous work in that it works in monochromatic extinctions rather than extinctions over a given filter bandpass. 
Hence integrating through the \cite{vergely2022} extinction cubes yields the extinction at 550\,nm (only roughly equivalent to $A_V$) which we then converted to extinction at a given wavelength using the standard Galactic value of $R(55)$ (again the rough equivalent $R_V$) of 3.02. 
Four objects (RZ~Gru, VY~Scl, VZ~Scl and DW~UMa) are all sufficiently far from the Galactic Plane that they lie above or below the $\pm$400\,pc thickness of the \cite{vergely2022} models, but given the small scale height of the Galactic dust layer \citep[around 100\,pc, e.g.][]{jones2011} and their small extinctions (0.04 mags) integrating out to the edge of the model should be sufficient for our purposes.

\begin{landscape}
\begin{table}
\begin{center}
\caption{\label{tab:system_par_dne} List of DNe analysed in this work. The table includes the object names, revised extinction values, the representative SWP and LWR/LWP spectrum spectra, inclination, DNe subtype, distance, orbital period, FES magnitude, outburst state, the time since the start of the outburst (where known), and the brightness state of the source at the time the spectra were taken.}
\begin{tabular}{lllllllllllll}
\hline\hline
Object    & $A_{55}$ & SWP    & LWR   & LWP   & Inclination & Subtype & Distance [pc] & $P\,(\textrm d)$ & FES & Outburst state & $\Delta t\,(\textrm d)$ & Brightness state\\
\hline
RX And    & 0.054 & 17673 & 13931 & -     & 51.0$\pm$9$^{\circ}$  & ZC  & 196.60 & 0.209893    & 11.1                 & Peak outburst                                                      & 0                                                       & Peak brightness  \\
AR And    & 0.104 & 18877 & 14885 & -     & 40.0$\pm$10$^{\circ}$ & UG  & 409.59 & -           & 12.4                 & Early decline                                                      & 1                                                       & Peak$-0.5$ \,mag \\
VY Aqr    & 0.043 & 21720 & -     & 02367 & 41.0$^{\circ}$        & WZ  & 141.13 & -           & 11.6                 & Decline                                                            & 6                                                       & Peak$- 1$\,mag   \\
Z Cam     & 0.028 & 18630 & 14697 & -     & 57.0$\pm$11$^{\circ}$ & ZC  & 213.50 & 0.289841    & 10.0                   & Peak / early decline                                               & 1                                                       & Peak$-0.5$\,mag  \\
SY Cnc    & 0.055 & 08082 & 07051 & -     & 56.7$\pm$20$^{\circ}$ & ZC  & 401.25 & 0.382375    & 11.5                 & Decline                                                            & 6                                                       & Peak$-0.5$\,mag  \\
YZ Cnc    & 0.040 & 03727 & 03308 &       & 38.0$\pm$3$^{\circ}$  & SU  & 233.49 & 0.0868      & 11.9                 & Early decline                                                      & 6                                                       & Peak$-$1\,mag    \\
HL CMa    & 0.175 & 19873 & 15857 & -     & 45.0$\pm$10$^{\circ}$ & ZC  & 293.54 & 0.216787    & 12.0                   & Outburst                                                           & -                                                       & -                \\
WW Cet    & 0.044 & 10664 & 09373 & -     & 54.0$\pm$9$^{\circ}$  & ZC  & 218.06 & 0.1758      & 12.2                 & Outburst                                                           & -                                                       & -                \\
SS Cyg    & 0.050 & 01788 & 01675 & -     & 37.0$\pm$5$^{\circ}$  & UG & 112.35 & 0.27513     & 8.6                  & Wide outburst                                                      & 5\footnote{Number of days since rise.} & Peak brightness  \\
AB Dra    & 0.253 & 17619 & 13886 & -     & 40.0$\pm$10$^{\circ}$ & UG  & 396.61 & -           & 12.7                 & Late rise                                                          & -1                                                      & Peak$-0.5$\,mag  \\
VW Hyi    & 0.019 & 24307 & -     & 04652 & 60.0$\pm$3$^{\circ}$  & SU  & 53.74  & 0.074271038 & 9.0                    & Super outburst                                                     & 2\footnote{Number of days since rise.} & Peak brightness  \\
WX Hyi    & 0.036 & 07509 & 06500 & -     & 40.0$\pm$10$^{\circ}$ & SU  & 228.42 & 0.074813    & 11.6                 & Super outburst                                                     & 0                                                       & Peak brightness  \\
X Leo     & 0.072 & 15951 & 12282 & -     & 41.0$\pm$10$^{\circ}$ & UG  & 438.11 & -           & 12.6                 & Outburst                                                           & -                                                       & -                \\
AY Lyr    & 0.088 & 09342 & 08098 & -     & 41.0$\pm$10$^{\circ}$ & SU  & 415.55 & -           & 13.2                 & Super outburst\footnote{With precursor outburst.} & 5                                                       & Peak brightness  \\
CY Lyr    & 0.307 & 21030 & 16779 & -     & 60.0$\pm$10$^{\circ}$ & UG  & 468.33 & -           & 13.2                 & Peak outburst                                                      & 2                                                       & Peak brightness  \\
CZ Ori    & 0.165 & 07358 & 06376 & -     & 18.0$\pm$10$^{\circ}$ & UG  & 499.97 & 0.2189      & 12.4                 & Decline                                                            & 4                                                       & Peak$-0.5$\,mag  \\
RU Peg    & 0.105 & 15062 & 11594 & -     & 41.0$\pm$7$^{\circ}$  & UG  & 271.30 & 0.3746      & 11.2                 & Mid-decline                                                        & 7                                                       & Peak$-1.5$\, mag \\
TZ Per    & 0.629 & 17643 & 13907 & -     & 56.7$\pm$20$^{\circ}$ & ZC  & 456.95 & -           & 12.9                 & Outburst                                                           & 6                                                       & Peak brightness  \\
KT Per    & 0.287 & 17712 & 13968 & -     & 60.0$\pm$10$^{\circ}$ & UG  & 244.66 & 0.162656    & 12.5                 & Peak outburst                                                      & 0                                                       & Peak brightness  \\
UZ Ser    & 0.601 & 17633 & 13901 & -     & 50.0$\pm$10$^{\circ}$ & ZC  & 308.83 & -           & 12.7                 & Outburst                                                           & -                                                       & -                \\
VW Vul    & 0.425 & 18875 & 14883 & -     & 41.0$\pm$10$^{\circ}$ & ZC  & 542.92 & -           & 13.9                 & Outburst                                                           & -                                                       & -                \\
AH Her    & 0.096 & 41939 & -     & 20704 & 46.0$\pm$3$^{\circ}$  & ZC  & 327.91 & 0.258116    & 12.3                 & Early decline\footnote{Outburst with slow rise.}  & 3                                                       & Peak$-0.8$\, mag \\
DX And    & 0.312 & 37689 & -     & 16844 & 45.0$\pm$12$^{\circ}$ & UG  & 585.37 & 0.440502    & 12.6$^{\ast}$        & Wide outburst                                                      & -1                                                      & Peak$-0.3$\,mag  \\
IR Gem    & 0.054 & 38524 & -     & 17696 & 15.0$\pm$10$^{\circ}$ & SU  & 257.83 & 0.0684      & 12.3                 & Peak outburst                                                      & 3                                                       & Peak brightness  \\
SU UMa    & 0.040 & 34824 & -     & 14532 & 43.0$\pm$10$^{\circ}$ & SU  & 220.14 & 0.07635     & 12.8                 & Early decline                                                      & <1                           & Peak$-0.5$\,mag  \\
TT Crt    & 0.077 & 47801 & -     & 25677 & 60.0$\pm$10$^{\circ}$ & UG  & 532.41 & 0.26842     & \dots & Early decline?                                                     & -                                                       & -                \\
V1159 Ori & 0.303 & 56781 & -     & 31957 & 35.0$\pm$10$^{\circ}$ & SU  & 351.17 & 0.062178    & \dots & Outburst                                                           & 11                                                      & Peak brightness  \\
V426 Oph\footnote{V426 Oph was classified as a NL in \cite{ladous1991}, but according to more recent catalogues \citep[e.g.][]{ritter2003} based on more recent observations, this object is a DN.}
  & 0.243 & 29546 & -     & 09427 & 59.0$\pm$6$^{\circ}$  & ZC  & 190.01 & 0.2853      & 11.7                 & Outburst                                                           & -                                                       & Peak brightness 
\\
\hline
\end{tabular}
\end{center}
\footnotesize{\textbf{Notes.} The distances are the geometric distances from the \protect\cite{bailerjones2021} catalogue based on Gaia measurements. The FES magnitudes flagged with an asterisk indicate that the brightness estimated from the FES counts differs by more than 0.5\,mag from the brightness suggested by the AAVSO light curve. Where an FES magnitude is not available this is indicated with an ellipsis. The DNe subtypes are: U Gem (UG), SU UMa (SU), Z Cam (ZC), WZ Sagittae (WZ).} \\
\footnotesize{$^{7}$Number of days since rise.} \\
\footnotesize{$^{8}$Number of days since rise.} \\
\footnotesize{$^{9}$With precursor outburst.} \\
\footnotesize{$^{10}$Outburst with slow rise.} \\
\footnotesize{$^{11}$V426 Oph was classified as a NL in \cite{ladous1991}, but according to more recent catalogues \citep[e.g.][]{ritter2003} based on more recent observations, this object is a DN.} \\
\end{table}
\end{landscape}

\begin{table*}
\begin{center}
\caption{\label{tab:system_par_nl} List of NLs analysed in this work. The table includes the object names, revised extinction values, the representative SWP and LWR/LWP spectra, inclination, NL subtype, distance, orbital period, and the FES magnitude.}
\begin{tabular}{llllllllll}
\hline\hline
Object    & $A_{55}$ & SWP    & LWR   & LWP   & Inclination & Subtype & Distance [pc] & $P\,(\textrm d)$ & FES \\
\hline
HL Aqr & 0.199 & 23325 & -     & 03647 & 18.0$^{\circ}$          & UX & 567.09 & 0.13557  & 13.4 \\
TT Ari & 0.152 & 06278 & 05448 &       & 30.0$\pm$10$^{\circ}$   & VY & 247.02 & 0.13755  & 10.9 \\
KR Aur & 0.213 & 14734 & 11299 & -     & 38.0$^{\circ}$          & VY & 458.89 & 0.162772 & 13.1 \\
KQ Mon & 0.105 & 16267 & 12516 & -     & \textless{}60$^{\circ}$ & UX & 612.67 & -        & 12.6 \\
VY Scl & 0.042 & 32594 & -     & 12365 & 30.0$^{\circ}$          & VY & 595.61 & 0.2323   & 12.7 \\
RW Sex & 0.060 & 07500 & 06494 & -     & 34.0$\pm$6$^{\circ}$    & UX & 221.63 & 0.245145 & 10.6 \\
IX Vel & 0.027 & 18578 & 14655 & -     & 57.0$\pm$2$^{\circ}$    & UX & 90.31  & 0.193927 & 9.3  \\
BZ Cam & 0.214 & 21251 & 16940 & -     & 15.0$^{\circ}$          & VY & 369.64 & 0.15353  & 12.6 \\
\hline
\end{tabular}
\end{center}
\footnotesize{\textbf{Notes.} The distances are the geometric distances from the \cite{bailerjones2021} catalogue based on Gaia measurements. Where an FES magnitude is not available this is indicated with an ellipsis. The NL subtypes are UX UMa (UX) and VY Scl (VY). Novalikes may also be classified based on analysis of the emission lines in their spectra; from the above list HL Aqr, TT Ari, and KQ Mon have been classified as also belonging to the SW Sex (SW) subtype (see \citealt{mizusawa2010}, \citealt{ritter2003} and \citealt{wolfe2013} respectively).}
\end{table*}

\begin{figure}
    \centering
    \includegraphics[width=\columnwidth]{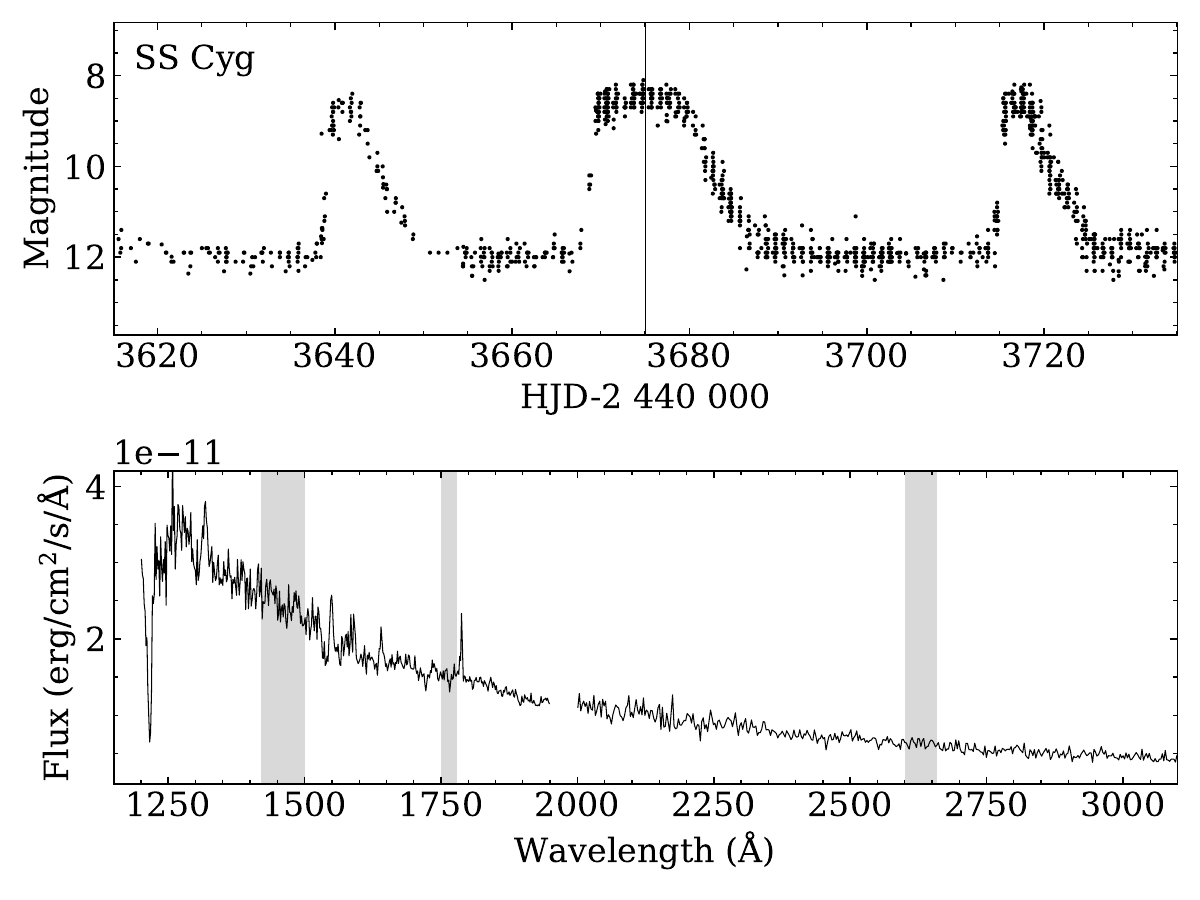}
    \caption{The light curve and the spectra of SS~Cyg. The top panel shows the AAVSO light curve with the time of the spectroscopic observation indicated with a vertical line. The bottom panel shows the \IUE\ SWP and LWR spectra of SS~Cyg. The grey shaded regions indicate the bands that we chose to obtain $C_{\textrm{short}}$ and $C_{\textrm{long}}$.}
    \label{fig:bands}
\end{figure}

\subsection{Final sample}

After collecting the NEWSIPS \IUE\ spectra from MAST for the extended sample, we applied the corrections determined by \cite{massa2000}, and we dereddened each spectrum using the \cite{fitzpatrick2019} extinction law. 
All sources, CV class and subclass,\footnote{We provide the subclasses based on the \cite{ritter2003} catalogue, which was last updated in 2011.} 
\IUE\ spectrum ID, extinction values, inclination, Gaia distances, orbital periods, and the FES magnitudes are listed in Tables~\ref{tab:system_par_dne} and \ref{tab:system_par_nl}. 
The FES magnitudes were calculated primarily based on the FES counts on the target (`$\textrm{CTS}$'), which is included in the headers of the SWP fits files. When either the $\textrm{CTS}$ or the `$\textrm{focus step}$' was not available in the SWP files, we estimated the FES magnitude from the LWR or the LWP files. 
We note that the FES magnitudes provide only an estimate on the system brightness. 
We indicated with asterisks those FES magnitudes, which differ by more than 0.5\,mag from the brightness estimated from the AAVSO light curves.

\section{Results}
\label{sect:results}

\subsection{Colours -- definition}
\label{sec:coldef}

To study and compare the UV spectra of DNe and NLs, we analysed the dereddened spectra in the following way. We first smoothed them by calculating the mean flux in a 9 pixel wide sliding window (box filter). 
We then chose three bands centred on $1460\,\AA$ with a width of $80\,\AA$, $1765\,\AA$ with a width of $30\,\AA$, and $2630\,\AA$ with a width of $60\,\AA$. 
We chose these central wavelengths and band widths so that we can minimize the contamination of any absorption or emission lines in the spectra, and so that they are also far enough from the edges where the spectra tend to be noisier. 
We chose two bands in the SWP spectra, because the effect we want to examine is expected to be more significant at shorter wavelength. 
We show an example spectrum with the selected bands in the bottom panel of Fig.~\ref{fig:bands}. 
We integrated the flux in each band for all of the objects in our sample. 
As a final step, we defined a short and a long wavelength colour by taking the negative logarithm of the ratio of the neighbouring bands, i.e.: 
\begin{equation}
C_{\rm short} = - \log \left( \frac{ \int_{\rm \lambda = 1420 \AA}^{1500 \AA} f_{\rm \lambda} \,\rm d \lambda}{\int_{\rm \lambda = 1750 \AA}^{1780 \AA} f_{\rm \lambda} \,\rm d \lambda} \right) =  - \log \left( \frac{F_{1460}}{F_{1765}} \right)
\end{equation}
and
\begin{equation}
C_{\rm long} = - \log \left( \frac{ \int_{\rm \lambda =1750 \AA}^{1780 \AA} f_{\rm \lambda} \,\rm d \lambda}{\int_{\rm \lambda =2615 \AA}^{2645 \AA} f_{\rm \lambda} \,\rm d \lambda} \right) = - \log \left( \frac{F_{1765}}{F_{2630}} \right).
\end{equation}

\begin{figure}
   \centering
   \includegraphics[width=\columnwidth]{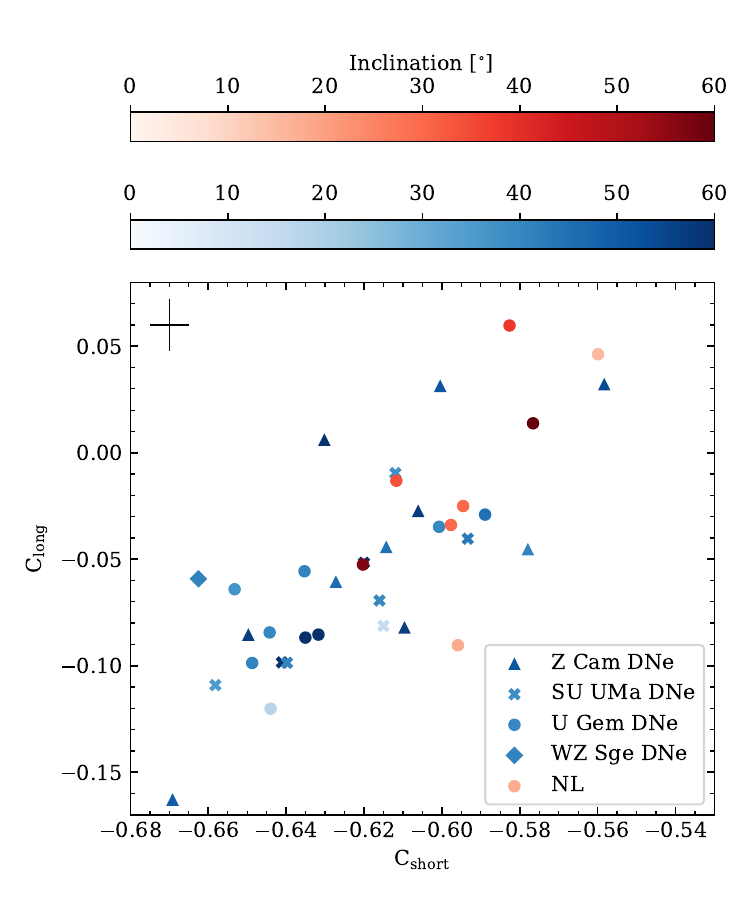}
   \caption{Colour-colour, $C_{\rm short}$ vs. $C_{\rm long}$, plot for the DNe and the NLs. The red symbols are the NLs. The blue symbols are the DNe, with triangles marking the Z Cam systems (which can show prolonged standstills), crosses marking the SU UMa systems (which can show superoutbursts) and circles marking the remaining DNe subtypes. 
   We have indicated the size of the representative errorbar with the black cross in the top left corner of the figure.
   As described in the text the dominant source of scatter in this plot is  intrinsic astrophysical variations of the sources which appears to be at the $\approx 0.02$ level in these quantities, with uncertainties in the data analysis at the level of $\approx 0.01$.
   The intensity of the colour indicates the inclination of the system as shown in the colour bar (see Tables~\ref{tab:system_par_dne} \& \ref{tab:system_par_nl} for details). This figure shows that the DNe and NLs are not drawn from the same parent population.}
   \label{fig:colour_types}
\end{figure}

The uncertainties in the flux ratios defined in this way are small.
Propagating the uncertainties from NEWSIPS through our analysis gives median uncertainties of 0.005 in $C_{\rm short}$ and 0.012 in $C_{\rm long}$, with a largest in either band being 0.018.   
These are comparable to, but smaller than, the differences we obtain between repeated measurements of the nova-likes IX~Vel and TT~Ari.  
The RMS between the $C_{\rm short}$ values of the 28 IX~Vel low-resolution large-aperture SWP spectra is 0.011. 
For the 37 high-state spectra of TT~Ari the RMS is 0.016.

\begin{figure}
   \centering
   \includegraphics[width=\columnwidth]{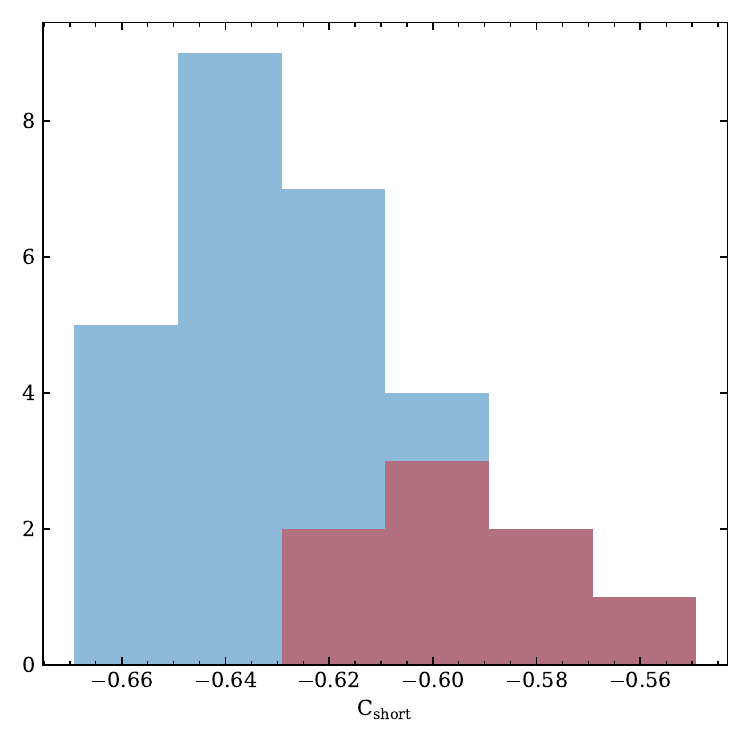}
   \caption{Histogram of the $C_{\textrm{short}}$ values for the DNe (blue bars) and NLs (red bars). It is clear that the two distributions are different.}
   \label{fig:hist_short}
\end{figure}

\begin{figure}
   \centering
   \includegraphics[width=\columnwidth]{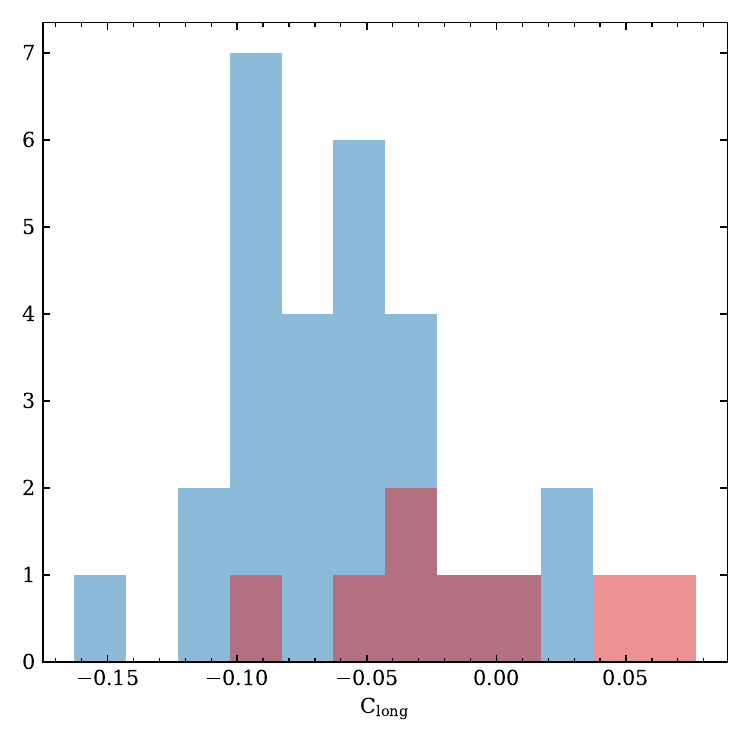}
   \caption{Histogram of the $C_{\textrm{long}}$ values for the DNe (blue bars) and NLs (red bars). Again it can be seen that the two distributions are distinct, but with less clarity than in Figure~\ref{fig:hist_short}.}
   \label{fig:hist_long}
\end{figure}

\subsection{Initial comparisons}
In Fig.~\ref{fig:colour_types}, we plot $C_{\rm short}$ against $C_{\rm long}$ in order to provide an initial comparison of the UV spectral slopes of the DNe and the NLs. 
The DNe and NLs are shown in different colours. 
It is immediately evident that the two sets of objects do not appear to have been drawn from the same parent population. 
To make this more apparent we compare in Fig.~\ref{fig:hist_short} histograms of the $C_{\rm short}$ for the DNe and the NLs, and in Fig.~\ref{fig:hist_long} the same for the $C_{\rm long}$. 
It is clear from the comparisons of the $C_{\rm short}$ distributions (Fig.~\ref{fig:hist_short}), and to a lesser extent the $C_{\rm long}$ distributions (Fig.~\ref{fig:hist_long}) that the UV spectra of the NLs in the high state are generally redder than the DNe in outburst.
Although it should be noted that there is some overlap in the two sets of distributions. 
Note, also, that the $C_{\rm short}$ and the $C_{\rm long}$ are not independent datasets.

Because of the paucity of data points (in particular for the NLs) it is not appropriate to make use of statistical tests (for example $\chi^2$) that rely on the assumption of underlying Gaussian distributions or approximations thereto (for example by application of the central limit theorem). 
Furthermore, as explained in Section \ref{sec:coldef}, the precise level of the uncertainties for individual datapoints is unclear.
We therefore make use of non-parametric statistics, and in particular the Kolmogorov-Smirnov (KS) test. 
Application of the 1D two sample KS test to the data in Fig.~\ref{fig:hist_short} ($C_{\rm short}$) shows that the probability that the null hypothesis (that both samples are drawn from the same parent population) being valid is less than or equal to 1.2 per cent. 
The same test applied to the data in Fig.~\ref{fig:hist_long} ($C_{\rm long}$) yields a probability of less than or equal to 3.8 per cent. 
Typically, the threshold for not rejecting the null hypothesis that the two samples are drawn from the same distribution is 5 per cent. 
It seems clear therefore that the conclusion of \cite{ladous1991}, that there is a significant difference between the UV spectra of DNe in outburst and of NLs in the high state, is upheld by our more detailed analysis which makes use of the most recent instrumental calibrations and also takes careful account of likely differential reddening between the DNe and the (on average) more distant NLs.

\begin{figure*}
   \centering
   \includegraphics[width=\textwidth]{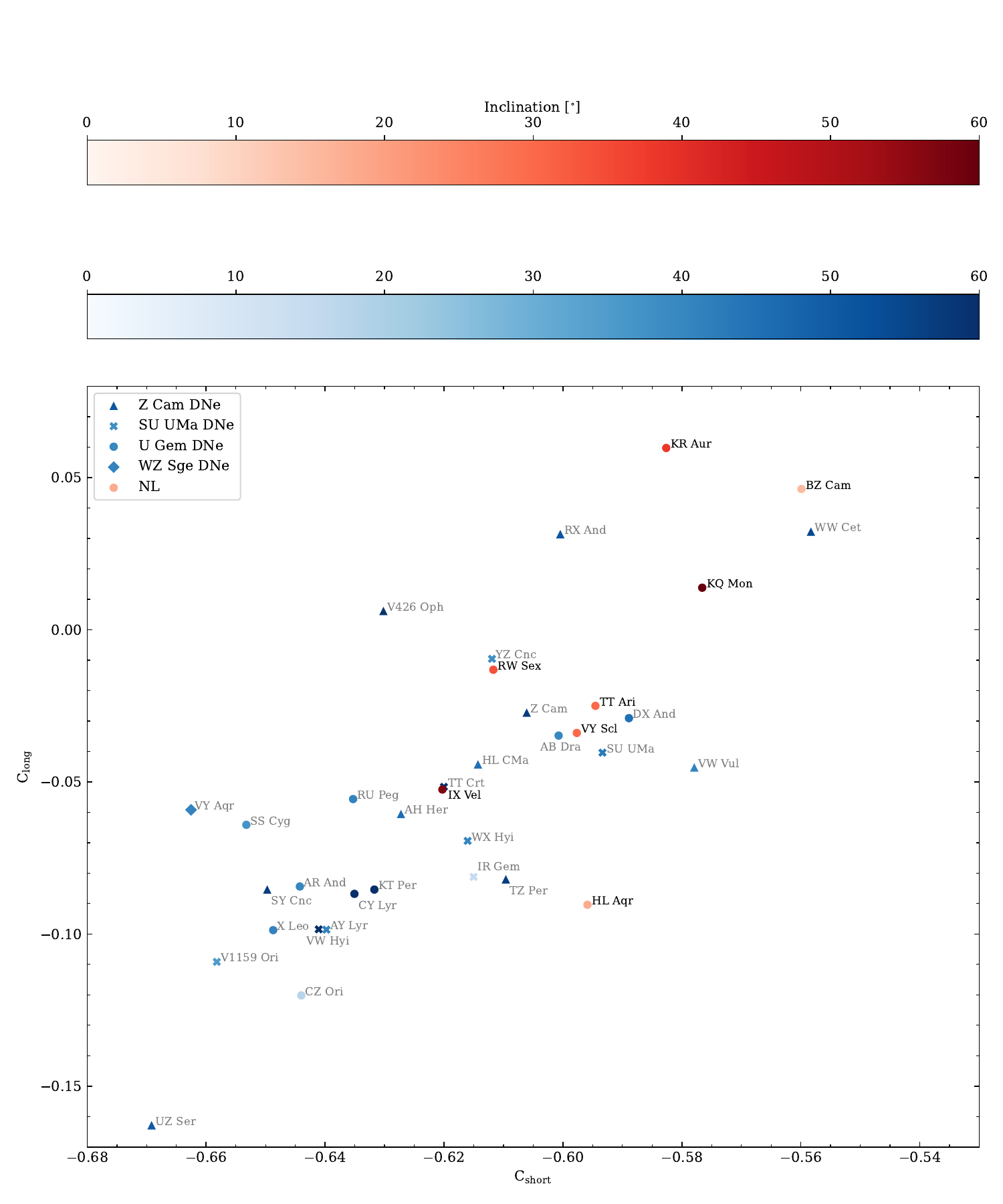}
   \caption{This figure shows the same plot as that in Fig.~\ref{fig:colour_types}, but here with the names of the systems included for reference.}
   \label{fig:colour_names}
\end{figure*}

\subsection{A more detailed comparison}

We here briefly look at the dataset in more detail. We construct a modified version of the $C_{\rm short}$ vs. $C_{\rm long}$ plot (Fig.~\ref{fig:colour_types}) in which we indicate the position of each object together with the name of the system (Fig.~\ref{fig:colour_names}). 
We can see from this plot that most of the DNe that were outliers (i.e. too red)  in the study of \cite{ladous1991} or were not fitted well (again largley because they were too red compared with the models) according to \cite{hamilton2007} are the ones whose positions lie in the region of overlap with the NL sample. 
These systems are the following: AB~Dra, WW~Cet, RX~And, VW~Vul, WX~Hyi, YZ~Cnc, Z~Cam, DX~And. 
There is also one source, WX~Hyi, which lies in a region of Fig.~\ref{fig:colour_names} that would suggest that good fit should be possible, but \cite{hamilton2007} find a poor fit. 
Comparing the \IUE\ spectra used by \cite{hamilton2007} (SWP23952) and the one used here (SWP07509) we find that the latter is at a significantly higher flux level (around a factor of 5 in the wavelength range $1420-1500$\,\AA) and thus we expect that the spectrum used by \cite{hamilton2007} for this source was not near the peak of the outburst.

\cite{puebla2007} modelled the UV spectra of a sample of CVs, which included 23 nova-like variables. 
They found that accretion-model spectra could not reproduce the observed spectra of the NLs, and concluded that almost all the models produce a bluer spectrum than the observations.
This was the case despite allowing
the system parameters ($M$, $\dot{M}$ and $i$) to be free parameters, which could lead to lower $\dot{M}$ and would mean that NLs also undergo disc instability. 
However, observations do not show disc instability in NLs. 
In addition they tried allowing the distance to be a free parameter in order to match the flux level of the observations.
It was the most poorly constrained parameter in their modelling procedure, and more accurate distance estimates now available from the \textit{Gaia} catalogue, reveal that the distance estimates from synthetic spectral fitting are lower by a factor of up to six than those measured by \textit{Gaia} \citep{gilmozzi2024}.
Our results, according to which the NLs spectra are redder than the DNe spectra, also suggest that the standard disc model will not reproduce the observed UV spectra of NLs.
This finding is in agreement with \cite{puebla2007}, who proposed that a revision of the temperature profile is needed to model the UV spectra more accurately, and this is particularly important for the innermost part of the disc.
Using the most up-to-date \textit{Gaia} distances will allow more accurate estimates of the system parameters, e.g. those estimated by \cite{gilmozzi2024}, and new efforts can be made to improve the modelling of the NL spectra.

\cite{godon2017} studied the entire \IUE\ CV sample, including both magnetic and non-magnetic sources. 
They determined the slope of the dereddened UV spectra using the wavelength range between ${\sim}1600\,$\AA\  and $3200\,$\AA, and measured the slope on a log-log scale. 
They found that the UV slope of the different disc-dominated CV subtypes decreases in the following order: SU~UMa, U~Gem, Z~Cam type DNe, then UX~UMa, and VY~Scl type NLs.
The location in Fig.~\ref{fig:colour_names} of the four UX~UMa type NLs (HL~Aqr, KQ~Mon, RW~Sex, IX~Vel) and the four VY~Scl type NLs (TT~Ari, KR~Aur, VY~Scl, BZ~Cam) in our sample reveals that our results are not inconsistent with the findings of \cite{godon2007}, though our sample is small. 
However, we must note that \cite{godon2007} studied the entire CV sample of \IUE\ (including systems with both low and high inclination), therefore, their results should not be discussed only in the context of intrinsic differences in accretion discs. 
They also note that the observed UV continuum emission of the SW~Sex type systems is probably affected by their high inclination.
In comparison, we focused on systems in which the disc dominates the UV spectrum (i.e. $i\leq60^{\circ}$), therefore, the differences we observe between the different CV classes and subclasses can be attributed to differences in their discs.

In Fig.~\ref{fig:colour_names} we have distinguished the different types of DNe. In particular we note here the ten Z~Cam type objects which are included in the analysis. 
These objects exhibit both DNe and NL-like characteristics; the continuous series of outbursts, which resemble those of standard DNe, are occasionally interrupted by the decline from outburst being halted by a long standstill, during which the the system brightness remains constant for a few weeks/months at a brightness level of around $0.5-1$ mag fainter that the peak outburst level. 
The standstills seem to be analogous to the behaviour of NLs which remain in a high state for long periods \citep[cf.][]{meyer1983}. 

Examining Fig.~\ref{fig:colour_names}, we can see that the Z~Cam stars whose colours overlap the majority of the other DNe are SY~Cnc, HL~CMa, TZ~Per, AH~Her and UZ~Ser (although we note that UZ Ser is the bluest source by a significant margin). 
From inspection of the AAVSO lightcurves (not depicted), we can be confident that all of these sources are near the maximum of DNe outbursts at the time the \IUE\ spectra were taken, with perhaps the exception of TZ~Per and HL~CMa for which short standstills cannot be ruled out. 
Z~Cam is on the boundary between the DNe group and the NL group, and from the AAVSO lightcurve appears to be in the DNe phase at the time the \IUE\ spectrum was taken. 
VW~Vul has a similar $C_{\rm long}$ to the DNe, but a larger $C_{\rm short}$; from the AAVSO lightcurve the state of VW~Vul is inconclusive, but may be during a standstill (although at a slightly higher flux level than previous standstills which may indicate a DNe state). 
V426~Oph has the opposite to VW~Vul with a similar $C_{\rm short}$ to the DNe and a similar $C_{\rm long}$ to the NLs; similar to other sources above the optical lightcurves indicate that V426~Oph was probably in a DNe outburst state, but there insufficient data to be definitive. 
The Z~Cam stars with colours most discrepant from the DNe group are WW~Cet and RX~And. 
For RX~And we note that there is another \IUE\ spectrum taken a day later that has an increased flux level (around 25 per cent higher) and appears bluer indicating that the spectrum we used may not have been near the peak of the outburst. 
For WW~Cet, the AAVSO lightcurve is inconclusive and it is possible that the source was in a standstill state. 

Given the possibility that some of the Z~Cam stars were not necessarily in the DNe outbursting state, we also performed the 1D KS test using a reduced DNe sample excluding the Z~Cam objects. 
This resulted in even smaller probabilities for the validity of the null hypothesis that the two samples (DNe and NLs) were drawn from the same parent population, being less than or equal to 0.3 per cent for $C_{\rm short}$ and less than or equal to 1.2 per cent for $C_{\rm long}$. 
This result highlights that Z~Cam stars may be an important class of stars for testing the evolution of the UV spectra with time, particularly through a prolonged standstill.

\section{Discussion}
\label{sect:discussion}

We have reanalysed the UV spectra obtained by \IUE\ for a sample of non-magnetic CVs comprising DNe in outburst and NLs that are sufficiently face-on to the line-of-sight that we have a good view of the accretion disc. 
Taking account of the NEWSIPS \IUE\ data processing, the \cite{massa2000} corrections, and updated distance and interstellar reddening estimates, we find that -- in agreement with the original analysis of \cite{ladous1991} and a more recent analysis by \cite{godon2017} -- the NLs have disc spectra that are significantly redder than the DNe. 
This is consistent with the extensive modelling efforts of the community that consistently find that (1) the outburst spectra of DNe can be fit by the standard disc model \citep{hamilton2007} and (2) the spectra of NLs cannot be fit by the standard model \citep{wade1988,long1994,linnell2007,linnell2007b,puebla2007,godon2017}. 
Our analysis substantiates that this difference cannot be accounted for by either interstellar reddening or statistical effects. 
We are therefore able to conclude that there is an intrinsic physical difference between these two very similar disc systems. 
Understanding this difference is likely to lead to a deeper understanding of the accretion process in general.

We note that, because the DNe in outburst are well-fit by the standard model, it is natural to assume that it is the NLs that are the ``outliers'' and must harbour some additional or alternative physical disc process. 
There is of course no guarantee that this assumption is correct. 
In this context we here draw attention to a couple of complementary ideas which may contain the germs of an explanation.

\cite{nixon2019} note that a physical difference between DNe in outburst and NLs is the time spent in the high accretion state \citep[see also][]{tout1993}. 
They present a model in which the disc, on entering the high accretion state, initially has only small-scale surface magnetic fields allowing for a local viscosity (i.e. the standard model). On timescales longer than the DNe outbursts, i.e. longer than at least a week or so, the lengthscale of the surface magnetic field structures grows, allowing for non-local transport of the accretion energy. 
\cite{nixon2019} suppose this occurs in the surface layers of the disc with mass transported along poloidal field lines to larger radii in what they call a Magnetically Controlled Zone (MCZ). 
They present simple calculations that show how this might affect the disc temperature profile, and are able to redistribute the energy in the disc in a way that may account for the difference in the observed spectra of DNe and NLs.

There have long been concerns about modelling accretion disc spectra \citep[see for example][]{pringle1992}, and in particular at what height in the disc the accretion energy is likely to be generated. 
The standard model, referred to above, makes the assumption that most of the accretion energy is deposited at high optical depth, on the grounds that most of the energy is likely to be deposited where most of the mass resides \citep{shakura1973}. 
However, this may well not be the case especially if the main viscous process is magnetic, since in that case the accretion energy appears in the form of magnetic energy which can then be dissipated elsewhere. 
Drawing on the ideas of \cite{kriz1986}, \cite{shaviv1986}, \cite{hubeny1990} and \cite{puebla2007} (see also the discussion by \citealt{pringle1992}),\footnote{Similarly, \cite{hassall1985} experiments with models that include an optically thin component finding that it can be a dominant component in the modelling of EK~TrA.} 
\cite{hubeny2021} provide models that can provide better fits to the NL spectra in which the accretion energy is deposited as a function of height through the disc, rather than all in the optically thick disc midplane region. 
This leads to emission from optically thin regions, which results in both a better representation of the UV slope and to filling in the Balmer jump, the depth of which is usually over-predicted in the models. 
The drawback with the modelling of \cite{hubeny2021} is that it does not provide any physical motivation for why the energy deposition would occur predominantly in the mid-plane in the DNe outbursts (as implied by the standard model fits) and predominantly at larger vertical heights in the NLs (their modelling of the NL IX~Vel finds that 99 per cent of the energy is dissipated at the disc surface).

It is plausible that the modelling of \cite{hubeny2021} provides a good description of the observational implications of the \cite{nixon2019} model. 
This might be expected because the \cite{nixon2019} model uses large scale magnetic field structures (in the NLs) to expel mass and deposit it at larger radii resulting in additional deposition of energy locally to the larger radii (the change in UV slope is effected by the lack of additional energy deposition at small radii in the disc for which there is no smaller radii to generate the surface mass flow). 
How the additional energy deposited by the surface flow is distributed through the vertical column of the disc depends on the mass flow rate in the MCZ and the surface density and temperature structure at the location at which mass is deposited. 
It is plausible that in some conditions the expelled matter landing on the disc at larger radii results in shocks that are confined to the surface layers, which could be commensurate with the model of \cite{hubeny2021} for IX~Vel.

We note that \cite{hartley2002} also proposes the presence of a large scale magnetic field in the NL discs, but they came to this conclusion by following a different approach. 
Their time-resolved spectroscopic observation of the nova-like variable IX~Vel do not reveal significant variability on short timescales (from $\sim$10 to $\sim$100\,s), and they conclude that the line-driven wind cannot be responsible for the observed level of mass-loss rate in this system. 
They propose that a large scale ordered magnetic field could be responsible for the observed level of mass-loss rate, if the streamline geometry required by the magnetic field controls the mass-loss rate in these systems. 
However, the large scale magnetic field that would be required to drive the wind need not to be the same as the one required for the MCZ, and we have not yet been able to address whether these are part of the same magnetic field structure.

\subsection{Investigating the time-dependence of the spectra from TT~Ari}
\label{sect:ttari}

If these ideas are in any way relevant, it would suggest that the differences found by \cite{ladous1991}, and substantiated here, between the UV spectra of DNe in outburst and NLs in the high state are caused by some difference in the magnetic structures in these discs, which thereby give rise to differences in where and how the accretion energy is deposited and turned into radiation. 
The obvious difference between the discs in outbursting DNe and those in NLs is the length of time that the discs have been in the high state \citep{tout1993,nixon2019}. 
In this picture the time taken for the disc to reach a steady state (the viscous timescale) is less than the time taken for the magnetic configuration to reach an equilibrium structure. 

Assuming the major difference in the spectra arises from the time spent in the high accretion state, and that this leads to magnetic structures as proposed by \cite{nixon2019}, then we can estimate the timescales we might expect in the following manner. 
The radius of a $0.6M_\odot$ white dwarf is $R_{\rm wd} \approx 0.01R_\odot$, and the orbital timescale at a radius of, say, $10R_{\rm wd}$ is $2\pi\sqrt{R^3/GM} \approx 300$\,s. 
The magnetic dynamo timescale is typically ten times the orbital timescale \citep[e.g.][]{tout1992,stone1996}, and so at $10R_{\rm wd}$ this is $\approx 3000$\,s, and so we expect the generation of small-scale fields with lengthscale $\sim H$ at the disc surface on timescales of order an hour. 
If there is a diffusive process to transfer magnetic field structures from size of order $H$ to size of order $R$, then the number of steps required in such a process is around $(R/H)^2$ \citep{king2004}, and here with $H/R \approx 0.03$ we have $(R/H)^2 \approx 10^3$. 
Thus, we estimate that the total time required to set up organised field on a scale of order $10R_{\rm wd}$ might be $\sim 3\times10^6$\,s, which is about a month. 
This is consistent with the split between DNe outbursts which last about a week and show standard disc spectra and the NLs which have been in the high accretion state for times much longer than a month and have spectra that require modification from the standard model.
Even VY~Scl type NLs, which spend some time in low accretion state, would have enough time to form a large scale magnetic field a few weeks/months after they returned to the high state. This is also supported by our result that all of the NL variables are redder than the average dwarf nova (Fig.~\ref{fig:colour_names}).
However, if we were to obtain UV spectra of a VY~Scl type system immediately after it has reached the maximum brightness, we might observe a bluer spectrum than a few months later.

\begin{figure}
  \centering
  \includegraphics[width=\columnwidth]{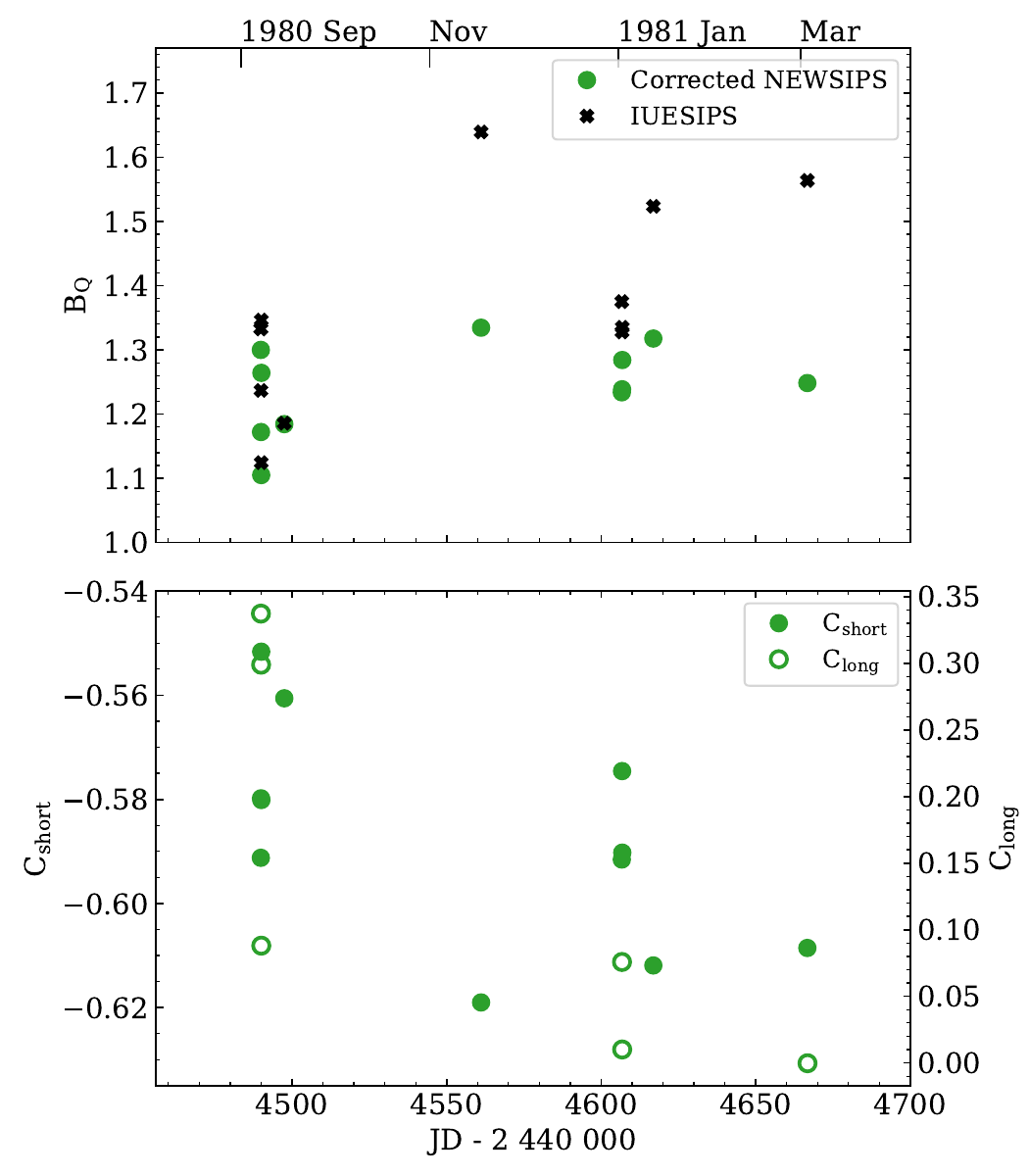}
  \caption{Top panel: The blueness parameter defined by \protect\cite{tout1993} for various TT Ari spectra, with black crosses corresponding to the IUESIPS data used by \protect\cite{tout1993} and green circles corresponding to the Corrected NEWSIPS data used here. The tick marks on the top axis show the beginning of each month. This figure shows that the re-analysis of the TT Ari data yields no significant change in the colour for these spectra. Bottom panel: The same as the top panel, but this time showing the $C_{\rm short}$ (filled circles) and $C_{\rm long}$ (open circles) UV colours we employ in this paper. 
  Again, there is no strong evidence for the evolution of the colours in these spectra.}
  \label{fig:ttari_blueness_compare}
\end{figure}

With such ideas in mind, \cite{tout1993} studied the \IUE\ spectra of TT~Ari, a VY~Scl type NL. 
These type of CVs spend extended time in a high state, which is occasionally interrupted by periods when the system brightness drops by several magnitudes. 
These low states may last several weeks to months. 
With the aim of examining the temporal evolution of a disc going through a cycle of high state and low state and then returning to the high state again, \cite{tout1993} measured the slope of the UV spectra by defining a `blueness' parameter ($B_{\rm Q}$) in the following way\footnote{Note that our $C_{\rm short}$ and $C_{\rm long}$ definitions are distinct, being both logged and inverted in comparison.},
\begin{equation}    
B_{\rm Q} = \frac{ \int_{\rm \lambda = 1318.25 \AA}^{1380.38 \AA} f_{\rm \lambda} \,\rm d \lambda}{\int_{\rm \lambda = 1819.7 \AA}^{1905.46 \AA} f_{\rm \lambda} \,\rm d \lambda}.
\end{equation}
Their results, which we have reproduced in the top panel of Fig.~\ref{fig:ttari_blueness_compare}, suggests that the UV spectrum of TT~Ari is `bluer', and therefore hotter, after the system returned to the high state from the brightness minimum (see black symbols in the top panel of Fig.~\ref{fig:ttari_blueness_compare}).

\begin{table}
\caption{Observing log of the TT~Ari spectra we used in Fig.~\ref{fig:ttari_blueness_compare}. }
\centering
\begin{tabular}{lll}
\hline\hline
SWP   & Date and time of observation       & $\textrm{JD}-2\,440\,000$ \\
\hline
10041 & 1980-09-07 09:27:17 & 4489.89      \\
10042 & 1980-09-07 10:54:22 & 4489.95      \\
10043 & 1980-09-07 12:17:58 & 4490.01      \\
10044 & 1980-09-07 13:36:28 & 4490.07      \\
10131 & 1980-09-14 23:38:56 & 4497.49      \\
10614 & 1980-11-17 15:25:37 & 4561.14      \\
10944 & 1981-01-02 04:00:50 & 4606.67      \\
10945 & 1981-01-02 06:21:50 & 4606.77      \\
10946 & 1981-01-02 07:39:05 & 4606.82      \\
11035 & 1981-01-12 08:40:37 & 4616.86      \\
13398 & 1981-03-03 04:33:59 & 4666.69      \\   
\hline
\end{tabular}
\label{tab:ttari_obslog}
\end{table}

However, with the aim of examining the impact of the new data reduction procedure on the scientific results, we re-analysed the IUESIPS low dispersion spectra used by \cite{tout1993} and employed the data analysis methodology outlined in Section~\ref{sect:observations} to produce the Corrected NEWSIPS low dispersion SWP spectra of TT~Ari. 
We summarised the log of observations in Table~\ref{tab:ttari_obslog}. 
As the bottom panel of Fig.~\ref{fig:ttari_spectra_compare} suggests, the new data reduction introduced changes in some of the spectra. Therefore, we calculated the $B_{\rm Q}$ blueness parameter as defined above using both the old and the new datasets. 
The results suggest, in contrast to the analysis of \cite{tout1993} on the IUESIPS data, that there was no significant change in the blueness parameter for the Corrected NEWSIPS data over the period analysed by \cite{tout1993}.

\begin{figure}
  \centering
  \includegraphics[width=\columnwidth]{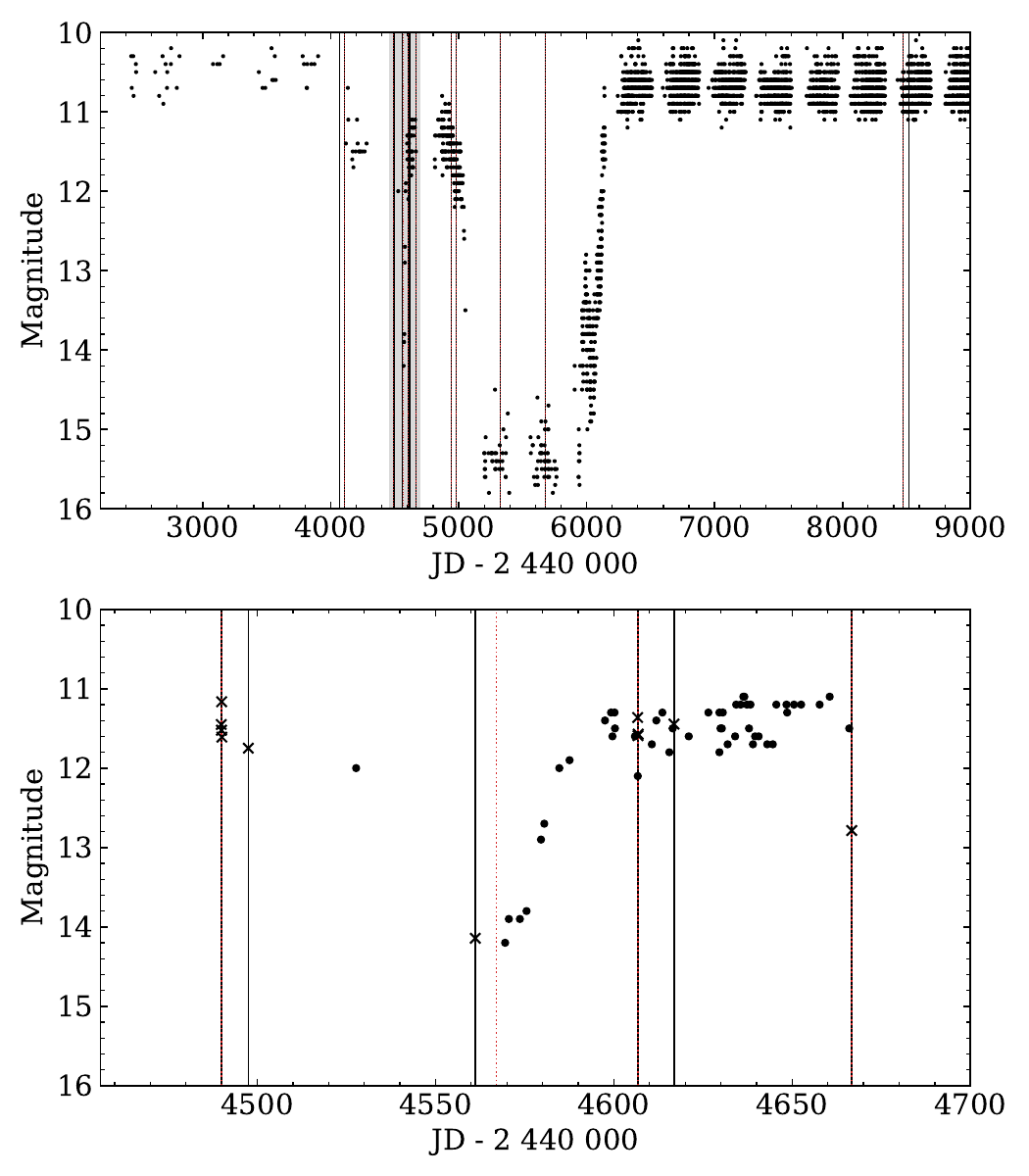}
  \caption{Top: The long-term optical (AAVSO) lightcurve of TT Ari for the entire \IUE\ observing period. The black vertical lines indicate the epochs at which SWP spectra were obtained, the red dotted lines mark the epochs of the LWP/LWR spectra. We plotted the red dotted line over the black solid line for the epochs when both SWP and LWP/LWR spectra were taken. The gray shaded area shows that region of the lightcurve which is enlarged in the bottom panel.  Bottom: The AAVSO lightcurve highlighting the time at which the spectra in Fig.~\ref{fig:ttari_blueness_compare} were taken. As can be seen in the figure the minimum prior to the time at which the spectra were taken is both shallow and short, and the recovery did not reach the long-term high state. As such it is clear that this was not a good minimum on which to attempt this analysis.}
  \label{TTAri_lightcurve}
\end{figure}

We must note, however, that the minimum \cite{tout1993} studied was very short and not well defined. 
Information on the system brightness from this period of time is available from the AAVSO database and we plot it in Fig.~\ref{TTAri_lightcurve}. 
These optical photometric data suggest that the system was indeed in a fainter state when some of the \IUE\ observations were carried out, but it is unclear from these data when the minimum started. 
Furthermore, this minimum was not as deep as other low states of TT~Ari. 
The FES magnitudes also confirm that the system was in a fainter state ($\sim$14\,mag) for one of the \IUE\ observations, but not as faint as during a regular low state ($\sim$16\,mag). 
Additionally, in the bottom panel of Fig.~\ref{fig:ttari_blueness_compare} we provide the $C_{\rm short}$ and $C_{\rm long}$ values for the Corrected NEWSIPS spectra over this period and find that vary in the range expected for the NLs based on the distributions in Fig.~\ref{fig:colour_types}. 
All in all, based on our re-analysis of the tentative results presented by \cite{tout1993} and our interpretation of the lightcurve of TT Ari around the time these spectra were taken, we can conclude that at present there is no significant data pointing towards an evolution of the colour of the spectra in these sources on a specific timescale. 
This question will need to be revisited in the future.

\section{Conclusions and Implications}
\label{sect:conclusion}

Through a reanalysis of the UV spectra of non-magnetic DNe in outburst and NLs from the \IUE\ satellite we have confirmed the findings of \cite{ladous1991}, and others, that the NL spectra are significantly redder than the DNe in outburst. 
This is an important discrepancy in our understanding of accretion discs. 
A physical explanation for this difference may generate significant understanding of the accretion process in a wide range of systems. 
We have provided discussion on the physical mechanisms that may be responsible and suggestions for future observations that may help to disentangle this mystery.

If, as we suggest, strong magnetic field structures are able to grow on the surfaces of the NL discs on the timescales indicated above, then this has significant implications for the behaviour of discs in other accreting sources and the application of the standard model to them. 
From the arguments in Section~\ref{sect:ttari}, the timescale to substantially modify the accretion process through growth of surface magnetic field structures is $t_{\rm MCZ} \sim (R/H)^2 t_{\rm dynamo}$, where again we shall take $t_{\rm dynamo} = ft_{\rm orb}$ with $f\approx 10$ and, as above, estimate the timescale at 10 stellar radii.\footnote{Note that the timescale here is $\propto (R/H)^2 t_{\rm orb}$ and scales in a way that is similar to the standard viscous timescale. So if the process is somehow not related to the disc dynamo, but instead nonetheless related to the viscous timescale, the numbers given here are still relevant.} 
We consider discs in X-ray binaries, AGN, and protostellar/protoplanetary discs.

The implications for X-ray binaries are clear as these discs are broadly similar to those in CVs, but with substantially reduced inner radii, and thus timescales in the inner regions, due to the more compact accretor. 
In this sense the inner regions of the disc are much ``older'' in terms of number of orbits than the CV discs. 
The disc thicknesses are also similar with $H/R \approx 0.01$. 
Here we have, as an illustration, a central black hole accretor of mass, say, $10M_\odot$ and thus gravitational radius $R_{\rm g} = GM/c^2 \approx 15$\,km. 
The orbital timescale at $10R_{\rm g}$ is therefore $\approx 0.01$\,s, and therefore the timescale $t_{\rm MCZ}$ is about a quarter of an hour.

For active galactic nuclei (AGN) the time- and length-scales are significantly larger, but the systems are also much longer lived. 
Scaling to a central black hole of mass $10^7M_\odot$ we have a gravitational radius of $\approx 0.1$\,au and an orbital timescale at $10R_{\rm g}$ of just under 3 hours. 
Employing the AGN structure derived by \cite{collin-souffrin1990}, from which we have $H/R \approx 0.002$ we get the timescale as $t_{\rm MCZ} \sim 10^3$\,yrs. 
As the lifetime of AGN discs is around $10^5-10^6$\,yrs \citep[e.g.][]{schawinski2015,king2015}, $t_{\rm MCZ}$ in the inner disc regions is only a small fraction of the typical disc lifetime.

For protoplanetary discs we have $H/R \approx 0.1$ \citep[e.g.][]{bell1997}, a central star of mass $\sim 0.5M_\odot$ and radius of, say, $2R_{\odot}$. 
In this case, the orbital timescale at 10 stellar radii is around 2 weeks. 
The dynamo timescale is then around half a year, and the timescale $t_{\rm MCZ}$ is tens of years. 
Again, this constitutes a small fraction of the $\sim$\,Myr lifetime of these discs. 
However, the ionisation structure of protoplanetary discs is complex, with large regions of the disc attaining sufficiently low levels of ionisation that magnetic activity can be shut off. 
We might therefore expect the effect to manifest most strongly in the innermost regions of protoplanetary discs, on scales $\lesssim 0.1-1$\,au, which are thought to remain sufficiently ionised by the radition field from the central star for magnetic activity to operate continuously \citep[e.g.][]{armitage2011}. 
Additionally, while protoplanetary discs are thought to spend most of their lifetimes in a low-viscosity state, at least some objects spend time in high viscosity states known as FU~Ori outbursts.
These can last longer than decades, and so may exhibit evolution in their surface magnetic structures similar to that inferred for the NLs.

In summary, if the dichotomy between the spectra of the DNe in outburst and the NLs is determined by the timescale for which these sources have been in the high accretion (fully ionised) state, and the critical timescale scales in the way we have described above, then almost all other discs in astronomy behave like the NLs rather than the DNe.\footnote{It is worth noting that the observations of Her X-1 provided by \cite{kosec2020} provide evidence for the gravitationally bound, surface outflows postulated by the \cite{nixon2019} model \citep[see][]{nixon2020}.} 
The conclusion of this line of thought is that almost all fully-ionised discs in the universe are {\it non-standard} in this manner. 
The discrepancy in spectra between DNe in outburst and NLs is therefore a significant unsolved question in our understanding of accretion discs, and addressing this question is likely to yield valuable information on the dynamical processes that govern disc evolution in a wide range of astrophysical systems.

Finally, we suggest that some useful next steps are:
\begin{enumerate}
\item To connect the \cite{nixon2019} and \cite{hubeny2021} modelling efforts to provide predictions for the time evolution of the UV spectra of CV discs upon (re)entering the high accretion state.
\item To obtain improved UV spectra, both in terms of quality and quantity, of both DNe in outburst and NLs in the high state to improve the statistics presented here.
\item And to catch discs that either enter a prolonged outburst (including the superoutbursts exhibited by SU UMa stars and the standstills exhibited by Z Cam stars) or that are recovering from a deep minimum (the VY Scl type stars) to explore the time-dependence of the UV spectra. 
\end{enumerate}
For example, {\it Hubble Space Telescope} spectra would be highly valuable as they would facilitate the analysis at shorter wavelengths than has been possible with the \IUE\ data here. 
From Figs.~9 \& 15 of \cite{linnell2007b} and Fig.~4 of \cite{nixon2019}, we anticipate that the effect is more pronounced at shorter wavelengths for which the emission originates from the central regions of the disc.


\section*{Acknowledgements}

GZs and CJN acknowledge support from the Leverhulme Trust (grant number RPG-2021-380). CJN acknowledges support from the Science and Technology Facilities Council (grant number ST/Y000544/1). This work has made use of data from the European Space Agency (ESA) mission Gaia (https://www.cosmos.esa.int/gaia), processed by the Gaia Data Processing and Analysis Consortium (DPAC, https://www.cosmos.esa.int/web/gaia/dpac/consortium). This research is based on observations made with the {\it International Ultraviolet Explorer}, obtained from the MAST data archive at the Space Telescope Science Institute, which is operated by the Association of Universities for Research in Astronomy, Inc., under NASA contract NAS 5–26555. We acknowledge with thanks the variable star observations from the AAVSO International Database contributed by observers worldwide and used in this research.

\section*{Data Availability}

The data underlying this article are available in the MAST and the AAVSO archives.




\bibliographystyle{mnras}
\bibliography{example} 





\bsp	
\label{lastpage}
\end{document}